\documentclass[conference]{IEEEtran}
\IEEEoverridecommandlockouts
\usepackage{cite}
\usepackage{amsmath,amssymb,amsfonts}
\usepackage{algorithmic}
\usepackage{graphicx}
\usepackage{textcomp}
\usepackage{xcolor}

\usepackage{multirow,setspace}
\usepackage{adjustbox} 
\usepackage{ulem}
\usepackage{soul}

\def\BibTeX{{\rm B\kern-.05em{\sc i\kern-.025em b}\kern-.08em
    T\kern-.1667em\lower.7ex\hbox{E}\kern-.125emX}}
\begin{document}

\title{Dynamic Graph Recommendation via Sparse Augmentation and Singular Adaptation
}

\author{\IEEEauthorblockN{Zhen Tao}
\IEEEauthorblockA{\textit{State Key Laboratory for Novel} \\
\textit{Software Technology}\\
\textit{Nanjing University}\\
Nanjing, China \\
zhentao@smail.nju.edu.cn}
\and
\IEEEauthorblockN{Yuehang Cao}
\IEEEauthorblockA{\textit{Laboratory for Big Data and Decision} \\
\textit{National University of Defense Technology}\\
Changsha, China \\
caoyuehang@nudt.edu.cn}
\and
\IEEEauthorblockN{Yang Fang}
\IEEEauthorblockA{\textit{National University of Defense Technology} \\
Changsha, China \\
fangyang12@nudt.edu.cn}
\and
\IEEEauthorblockN{Yunhui Liu}
\IEEEauthorblockA{\textit{State Key Laboratory for Novel} \\
\textit{Software Technology}\\
\textit{Nanjing University}\\
Nanjing, China \\
lyhcloudy1225@gmail.com}
\and
\IEEEauthorblockN{Xiang Zhao}
\IEEEauthorblockA{\textit{Laboratory for Big Data and Decision} \\
\textit{National University of Defense Technology}\\
Changsha, China \\
xiangzhao@nudt.edu.cn}
\and
\IEEEauthorblockN{Tieke He$^*$}
\IEEEauthorblockA{\textit{State Key Laboratory for Novel} \\
\textit{Software Technology}\\
\textit{Nanjing University}\\
Nanjing, China \\
hetieke@gmail.com}
\thanks{$^*$Corresponding author.}
}


\maketitle

\begin{abstract}
Dynamic recommendation, focusing on modeling user preference from historical interactions and providing recommendations on current time, plays a key role in many personalized services. Recent works show that pre-trained dynamic graph neural networks (GNNs) can achieve excellent performance. However, existing methods by fine-tuning node representations at large scales demand significant computational resources. Additionally, the long-tail distribution of degrees leads to insufficient representations for nodes with sparse interactions, posing challenges for efficient fine-tuning. To address these issues, we introduce GraphSASA, a novel method for efficient fine-tuning in dynamic recommendation systems. GraphSASA employs test-time augmentation by leveraging the similarity of node representation distributions during hierarchical graph aggregation, which enhances node representations. Then it applies singular value decomposition, freezing the original vector matrix while focusing fine-tuning on the derived singular value matrices, which reduces the parameter burden of fine-tuning and improves the fine-tuning adaptability. Experimental results demonstrate that our method achieves state-of-the-art performance on three large-scale datasets.
\end{abstract}


\section{Introduction}
Dynamic Recommendation Systems (DRS) are ingeniously designed to analyze and process real-time data, offering users personalized recommendations that are tailored to their immediate needs and preferences~\cite{ref1,ref2}.
Early dynamic recommendation systems involve time-sensitive sequential recommendation, which can be broadly divided into three categories, attention mechanisms-based~\cite{ref3,ref4}, graph neural networks-based~\cite{ref5,ref6} and contrastive learning-based methods~\cite{ref7,ref8}.
These methods achieve certain results in dynamic recommendation, but they mostly rely on static historical data and are hard to adapt to the ever-changing data streams.
Recently, Dynamic Graph Neural Networks (DGNNs) have been making significant strides in the realm of dynamic graph representation learning~\cite{ref9,ref10,ref11,ref12,ref13,ref_add1,ref_add2,ref_add3}. Moreover, the pre-training and fine-tuning paradigm has proven to be highly effective in graph models, adeptly extracting and transferring knowledge from pre-trained graphs to downstream tasks~\cite{ref14,ref15,ref16,ref17,ref18,ref19,ref20}. 
Consequently, the adoption of pre-training and fine-tuning in DGNNs has emerged as a dominant paradigm~\cite{ref21,ref22,ref23}.
Among them, a typical work is Graphpro~\cite{ref24}, which proposes a dynamic recommendation method by integrating temporal prompts with pre-training and fine-tuning, achieving promising results.

Despite the great success in dynamic recommendation by utilizing pre-trained DGNNs, there still remain two challenges:
(i) \textit{Low fine-tuning efficiency.} 
In dynamic recommendation, the existing methods usually adopt the form of pre-training and fine-tuning.
However, whenever making downstream recommendations based on the latest data, the model needs to undergo very costly fine-tuning.
Specifically, dynamic graph recommendations often require learning representations for a large number of nodes and then for downstream recommendations.
During this process, the graph changes temporally so that the representation of nodes needs to be updated to ensure the real-time performance of the model, which leads to substantial resource consumption for fine-tuning the node representations.
Therefore, there is an urgent demand for fine-tuning efficiency, which is not addressed by existing methods.
(ii) \textit{Ignoring node interaction sparsity.}
The probability distribution of node degrees in recommendation graphs typically follows a power-law distribution \cite{ref25,ref26}, necessitating high-dimensional vectors to effectively capture information for highly interactive nodes.
However, nodes with fewer interactions cannot effectively utilize the high-dimensional space\cite{ref27}, resulting in insufficient representations that are unfavorable for matrix decomposition and efficient fine-tuning.

To address the issues above, we propose a novel method \underline{Graph}-based \underline{S}parse \underline{A}ugmentation and \underline{S}ingular \underline{A}daptation (GraphSASA) for dynamic recommendation. 
Firstly, we present a singular adaptation fine-tuning technique, drawing inspiration from LoRA~\cite{ref28}, which is used in the transformers of LLMs. Our hypothesis is that the alterations in weights of node representations during the embedding fine-tuning process are characterized by a notably low "intrinsic rank". 
To harness this property, we apply Singular Value Decomposition (SVD) to the pre-trained enhancement matrix, meticulously extracting the top-r rank matrices. 
Subsequently, these matrices undergo a fine-tuning process, augmented by the elaborately designed dropout technique.
In this way, this approach greatly reduces the number of parameters that need to be fine-tuned and enables low dimensional singular matrices to proficiently capture and respond to temporal changes during the fine-tuning phase.
Secondly, we craft an innovative multi-level test-time augmentation approach for nodes with sparse interactions in graph convolutional networks. 
By selectively adding edges to low-degree nodes based on node representation similarities, we enhance their interaction and representational quality, leading to more robust fine-tuning.

In summary, the contributions are summarized as follows:
\begin{itemize}
\item We introduce an efficient fine-tuning method to node representations in pre-trained graph models and introduce the GraphSASA method, which reduces the number of fine-tuning parameters and better accommodates small temporal deviations. 
\item We propose a multi-level test-time augmentation technique that improves low-degree node representations to effectively cooperate with efficient fine-tuning. 
\item Experimental results demonstrate that GraphSASA achieves state-of-the-art performance on three large-scale dynamic recommendation datasets.
\end{itemize}
\begin{figure*}[h]
    \centering
    \includegraphics[width=\linewidth]{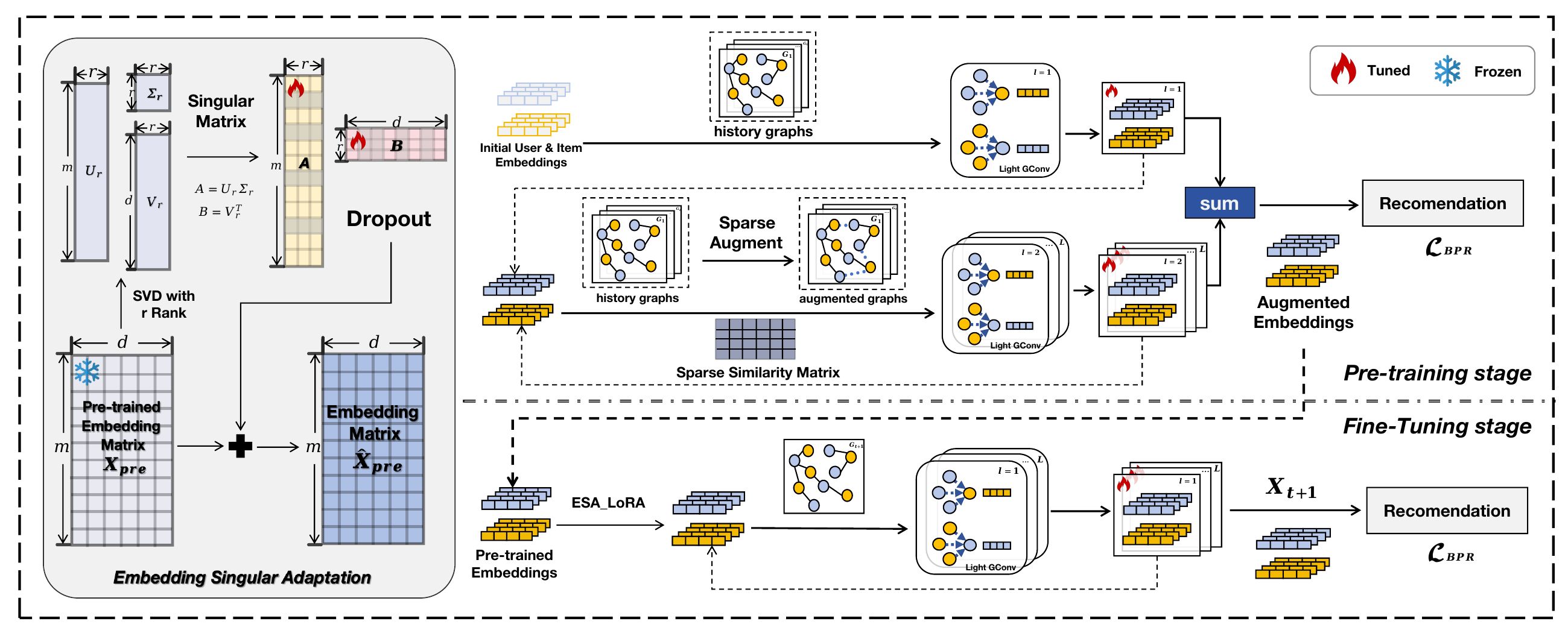}
    \caption{The overview of the proposed GraphSASA method.}
    \label{fig:enter-label}
\end{figure*}
\section{METHODOLOGY}
We provide an overview of our GraphSASA method, as illustrated in Fig. 1. The initial user and item embeddings undergo enhanced aggregation by adding augmented edges to low-degree nodes during the aggregation process, resulting in augmented embeddings. The pre-trained embedding matrix generated during pre-training is then efficiently fine-tuned through Embedding Singular Adaptation to produce the final node representations.
\subsection{Problem Definition}
\noindent\textbf{Graph and Nodes.} We construct a user-item interaction graph \( \mathcal{G} = (\mathcal{V}, \mathcal{E}) \), where the node set \( \mathcal{V} = \mathcal{U} \cup \mathcal{I} \) consists of users and items, with \( m \) total nodes, \( n_u \) users, and \( n_i \) items. The edge set \( \mathcal{E} \) represents interactions between users and items. The adjacency matrix is \( \mathcal{A} \in \mathbb{R}^{m \times m} \).

\noindent\textbf{Dynamic Graph Recommendation.} In dynamic environments, the user-item interaction graph evolves over time and is represented as a series of snapshots \(\{\mathcal{G}_1, \mathcal{G}_2, \ldots, \mathcal{G}_t\}\). Each snapshot \( \mathcal{G}_t = (\mathcal{V}_t, \mathcal{E}_t)\) contains a node set \( \mathcal{V}_t \subseteq \mathcal{V} \) and interaction edges \( \mathcal{E}_t \subseteq \mathcal{E} \) within a specific time interval \([T_{t-1}, T_t]\). 
We utilize LightGCN~\cite{ref29} with temporal convolutional~\cite{ref24} as the Backbone model.
\subsection{Sparse Augmentation}
\label{ssec:subhead}
To fully exploit the semantic information in the snapshot graphs, we apply multi-level test-time augmentation (TTA) to \( \mathcal{G}_t \) for nodes with sparse interactions.
Within our framework, the vector representations at time \( t-1 \), denoted as \( X_{t-1} \), is combined with the interaction graph  \( \mathcal{G}_t \), to learn the vector representations:
\begin{equation}
{X}_{t}=\mathrm{Backbone}({X}_{t-1},\mathcal{G}_{t}),
\end{equation}
where \(X_t\) is composed of graph convolution vectors. Let \( {h}^{(l)} \in \mathbb{R}^{(n_{u}+n_{i})\times d} \) denote the vector in the \( l \)-th layer of the graph convolution, where \( d \) represents the dimension.
The set of low-degree user nodes, \( \mathcal{U}_{\mathrm{low}} \), is selected based on the degree threshold \( \theta \): \( \mathcal{U}_{\mathrm{low}}=\{u \in \mathcal{U} \mid \deg(u) \leq \theta\} \). We compute the similarity matrix \( \mathcal{S} \in \mathbb{R}^{|\mathcal{U}_{\mathrm{low}}| \times |\mathcal{I}|} \) between embeddings of \( u \in \mathcal{U}_{\mathrm{low}} \) and \( i \in \mathcal{I} \) using the cosine distance (or alternatively, the euclidean distance):
\begin{equation}
\begin{aligned}s_{(u,i)}=&\phi(h_{u},h_{i})\\=&\frac{h_{u}\cdot h_{i}}{\|h_{u}\|\cdot\|h_{i}\|},\end{aligned}
\end{equation}
where \(\|\cdot\| \) denotes vector norm. From \( \mathcal{S} \), we select \( k \) similar items \( i \in \mathcal{I} \) for each \( u \in \mathcal{U}_{\mathrm{low}} \) to form \( k \) augmented edges, which are then added to the original graph \( \mathcal{G} \).
\begin{equation}
\mathcal{A}_{u,i}^{\text{aug}}=\begin{cases}1, & s_{u,i} \in \text{top-}K(\{s_{u,j}, j \in \mathcal{I}\}) \\ 0, & \text{otherwise}.\end{cases}
\end{equation}

The adjacency matrix of the augmented graph $\mathcal{G}'$ is defined as $\mathcal{A}^{\prime} = \mathcal{A} \cup \mathcal{A}^{\text{aug}}$. We perform edge augmentation with a probability of \( \lambda \) at each layer of graph convolution to enhance the model's flexibility and generalization capabilities:
\begin{equation}
h^{(l+1)} = \begin{cases} \operatorname{GConv}(h^{(l)}, \mathcal{G}_t) \\ \operatorname{GConv}(h^{(l)}, \mathcal{G}'_t). \end{cases}
\end{equation}

The final representation is obtained by summing the embeddings \( h \), which aggregate the semantic information from different layers.
\begin{equation}
{h}_{\text{final}} = \sum_{l=1}^L {h}^{(l)}.
\end{equation}
\subsection{Singular Adaptation}
\label{ssec:subhead}
After pre-training, we obtain the vector representation matrix and define it as \( X_{pre} \), which includes \( X_{user} \) and \( X_{item} \). The original matrix \( X_{pre} \) is then fine-tuned using singular adaptation. Firstly, we load the pre-trained matrix \( {X}_{pre} \in \mathbb{R}^{m \times d} \) and perform SVD on it:
\begin{equation}
{X}_{pre} = U S V^T,
\end{equation}
where \( U \in \mathbb{R}^{m \times m} \) is the left singular matrix, \( S \in \mathbb{R}^{m \times d} \) is the diagonal matrix of singular values, and \( V \in \mathbb{R}^{d \times d} \) is the right singular matrix. Since the first \( r \) singular values retain most of the information from the original matrix \cite{ref30}, we extract the first \( r \) singular values and their corresponding matrices 
\(
U_r = U[:, :r] \in \mathbb{R}^{m \times r}, S_r = S[:r, :r] \in \mathbb{R}^{r \times r},  V_r = V[:, :r] \in \mathbb{R}^{d \times r}
\), 
and then obtain the low-rank matrices 
\(
A = U_r S_r \in \mathbb{R}^{m \times r},  B = V_r^T \in \mathbb{R}^{r \times d}
\).
The pre-trained matrix \( X_{pre} \) can be approximated as:
\begin{equation}
X_{pre} \approx U_r S_r V_r^T = AB.
\end{equation}

Then, we design and apply the dropout method \cite{ref31,ref32} to the low-rank matrices, introducing random noise to enhance generalization and prevent overfitting.
\begin{equation}
\begin{aligned}
\hat{A}=\mathrm{diag}(m_A)  \cdot A,\quad m_A\sim\text{Bernoulli}(1-p),\\
\hat{B}=B\cdot\mathrm{diag}(m_B),\quad m_B\sim\text{Bernoulli}(1-p),\\
\end{aligned}
\end{equation}
where $p$ is the dropout probability, \( m_A \) and \( m_B \) are Bernoulli random vectors for matrices \( A \) and \( B \), respectively. \( \mathrm{diag}(m_A) \) randomly drops rows from \( A \), while \( \mathrm{diag}(m_B) \) randomly drops columns from \( B \). The vector representations is updated as:
\begin{equation}
\hat{X}_{pre} = X_{pre} + \hat{A} \hat{B}.
\end{equation}

During fine-tuning on the graph at time \( t+1 \), the parameters of \( X_{pre} \) are frozen, and only the low-rank matrices are updated:
\begin{equation}
X_{t+1} = \text{forward}(\hat{X}_{pre}, \mathcal{G}_{t+1}).
\end{equation}

The original embedding matrix \( X_{pre} \in \mathbb{R}^{m \times d} \) is decomposed into low-rank matrices \( A \in \mathbb{R}^{m \times r} \) and \( B \in \mathbb{R}^{r \times d} \) using singular value decomposition (SVD). The ratio of the number of parameters in the fine-tuned low-rank matrices to that in the original matrix can be expressed as:

\begin{equation}
\rho = \frac{(m+d) \times r}{m \times d} \approx \frac{r}{d}.
\end{equation}

Moreover, the time deviation between fine-tuning and the last historical graph is usually small, allowing the low-rank matrices to better fit temporal changes during fine-tuning.

\subsection{Optimization Objective and Loss Function}
\label{ssec:subhead}
The dot product of the final representations of the user and item, \(\hat{y}_{ui} = \mathbf{e}_u^T \mathbf{e}_i\), is used as the ranking score for generating recommendations. The model employs a loss function based on Bayesian Personalized Ranking (BPR) \cite{ref33}:
\begin{equation}
L = -\sum_{(u, i, j) \in D} \log \sigma(\hat{y}_{ui} - \hat{y}_{uj}),
\end{equation}
where \(\hat{y}_{ui}\) represents the predicted score for user \(u\) and the positive item \(i\), while \(\hat{y}_{uj}\) denotes the predicted score for the negative item \(j\), \(D\) is a dataset with negative sampling. The objective is to maximize the margin between the positive and negative sample scores.
\begin{table*}[t]
\setlength{\tabcolsep}{7.5pt}
\renewcommand{\arraystretch}{0.95}
\centering
\caption{Performance comparison of GraphSASA method with baseline methods on three datasets. The best performance is bolded, and the second best is underlined.
}
\begin{tabular}{lcccccc}
\hline
\multirow{2}*{\textbf{Method}} & \multicolumn{2}{c}{\textbf{Taobao}}  & \multicolumn{2}{c}{\textbf{Amazon}} & \multicolumn{2}{c}{\textbf{Koubei}}  \\
\cline{2-7}
 & {Recall@20}   & {nDCG@20}    & {Recall@20}   & {nDCG@20}    & {Recall@20}   & {nDCG@20}    \\
\hline
GraphPrompt        & 0.0199            & 0.0195           & 0.0154            & 0.0075           & 0.0342            & 0.0249           \\ 
GPF                & 0.0223            & 0.0220           & 0.0174            & 0.0088           & 0.0348            & 0.0251           \\ 
EvolveGCN-H        & 0.0224            & 0.0221           & 0.0138            & 0.0066           & 0.0315            & 0.0231           \\ 
EvolveGCN-O        & 0.0236            & 0.0232           & 0.0157            & 0.0084           & 0.0334            & 0.0242           \\ 
ROLAND             & 0.0226            & 0.0226           & 0.0150            & 0.0069           & 0.0301            & 0.0223           \\ 
DGCN               & 0.0229            & 0.0228           & 0.0158            & 0.0084           & 0.0353            & 0.0255           \\ 
GraphPro           & \underline{0.0251} & \underline{0.0245} & \underline{0.0191} & \underline{0.0094} & \underline{0.0362} & \underline{0.0265} \\ 
GraphSASA          & \textbf{0.0263}   & \textbf{0.0258}  & \textbf{0.0205}   & \textbf{0.0101}  & \textbf{0.0368}   & \textbf{0.0270}  \\
\hline
\end{tabular}
\end{table*}
\section{Evaluation}
\subsection{Experimental Setup}
\label{ssec:subhead}
\noindent\textbf{Datasets and Metircs.} We utilized three datasets that reflect real-world scenarios in dynamic recommendation, including Taobao dataset, Amazon dataset, and Koubei dataset. The snapshot intervals for these datasets are daily, weekly, and weekly, respectively. Following the datasets settings of GraphPro \cite{ref24}, we simulated dynamic changes using graph snapshots generated at different time intervals, starting with pre-training and subsequently fine-tuning the dynamic GNN. We employed the standard metrics, Recall@k and nDCG@k, as our evaluation metrics. Specifically, we set $k = 20$ and averaged the results over all future snapshots.

\noindent\textbf{Baseline Methods.} We compare our model with three types of methods. Dynamic recommendation methods include DGCN \cite{ref9} and GraphPro \cite{ref24}. Dynamic graph neural network methods include EvolveGCN-H \cite{ref11}, EvolveGCN-O \cite{ref11} and ROLAND \cite{ref12}. Graph prompting methods include GraphPrompt \cite{ref16} and GPF \cite{ref17}.

\noindent\textbf{Implementation Details.} The threshold $k$ and $\lambda$ are searched from \{1,\,3,\,5\} and \{0.8,\,0.9,\,0.95\}, respectively. Sparse augmentation with 0.001 for low-degree nodes. The rank $r$ is set to 16, and the embedding dimension $d$ is set to 64. 
\subsection{Baseline Performance Comparison}
\label{ssec:subhead}
The results presented in Table \uppercase\expandafter{\romannumeral1} show that the GraphSASA method outperforms all other methods across all datasets. Several factors contribute to this superior performance: First, GraphSASA employs a temporal convolution model, which, together with the next-best results \cite{ref24}, highlights its effectiveness in capturing time-varying characteristics in dynamic graphs. Second, the sparse augmentation in GraphSASA enhances dynamic graph node representations by leveraging the similarity of node representation distributions during the graph hierarchical aggregation process. Third, with snapshots taken at daily or weekly intervals, the low-rank singular value matrices used in the fine-tuning process effectively retain key structural features while better fitting smaller dynamic changes.

\subsection{Analysis of GraphSASA}
\label{ssec:subhead}
\noindent\textbf{Ablation Study.} As shown in Table \uppercase\expandafter{\romannumeral2}, we conducted ablation experiments on the SA (Sparse Augmentation) and ESA (Embedding Singular Adaptation) components. In results demonstrate that SA and ESA are effective in enhancing performance. The results of the ``w/o SA" (without SA) configuration indicate that the sparse augmentation strategy helps the model learn more comprehensive node representations from the snapshot graphs. The results of the ``w/o ESA" (without ESA) show that Embedding Singular Adaptation is effective in enhancing the model's performance. The combination of low-rank singular matrices with dropout allows the model to better adapt to minor temporal variations during the fine-tuning phase.
\begin{table}[h]
\centering
\caption{Ablation study on key components of GraphSASA.}
\renewcommand{\arraystretch}{1.15}
\begin{adjustbox}{width=0.97\linewidth}
\begin{tabular}{lcccc}
\hline
\multirow{2}*{\textbf{Method}}    & \multicolumn{2}{c}{\textbf{Taobao}}  & \multicolumn{2}{c}{\textbf{Amazon}} \\
\cline{2-3}
\cline{4-5}
                   & {Recall@20}   & {nDCG@20}    & {Recall@20}   & {nDCG@20}   \\
\hline
w/o SA             & 0.0255            & 0.0249           & 0.0195            & 0.0096          \\
w/o ESA            & 0.0258            & 0.0253           & 0.0199            & 0.0098          \\
GraphSASA          & \textbf{0.0263}   & \textbf{0.0258}  & \textbf{0.0205}   & \textbf{0.0101} \\
\hline
\end{tabular}
\end{adjustbox}
\end{table}

\begin{table}[h]
\centering
\caption{Comparison of memory usage and performance.}
\renewcommand{\arraystretch}{1.2}
\begin{adjustbox}{width=0.99\linewidth}
\begin{tabular}{lcccc}
\hline
\textbf{Dataset} & \textbf{Model}   & {Recall@20} & {nDCG@20}  & {Memory Usage} \\
\hline
\multirow{2}*{\textbf{Taobao}}           & GraphPro         & 0.0251          & 0.0245         & 6291MB \\
                  & GraphSASA        & 0.0263          & 0.0258         & 4098MB \\
\hline
\multirow{2}*{\textbf{Amazon}}           & GraphPro         & 0.0191          & 0.0094         & 3221MB \\
                  & GraphSASA        & 0.0205          & 0.0101         & 1799MB \\
\hline
\end{tabular}
\end{adjustbox}
\end{table}

\noindent\textbf{Memory Usage Analysis.} As shown in Table \uppercase\expandafter{\romannumeral3}, GraphSASA exhibits lower memory consumption during fine-tuning compared to GraphPro. The memory usage during fine-tuning includes the memory required for the fine-tuning matrices, as well as other variables within the model, such as the vectors used in aggregation and the edges in the graph. GraphSASA reduces memory usage by decreasing the number of parameters in the fine-tuning matrices through the use of singular matrices. Compared to GraphPro, GraphSASA achieves superior performance while reducing memory usage by approximately 34\% on the Taobao dataset and 44\% on the Amazon dataset. This indicates that GraphSASA offers a significant memory advantage for fine-tuning large-scale graph datasets.

\section{Conclusion}
\label{sec:page}
This paper presents a novel method named GraphSASA for efficient fine-tuning of dynamic recommendation. GraphSASA enhances vector representations by adding augmented edges to nodes with sparse interaction based on representation distribution similarity during the graph hierarchical aggregation process. By applying singular adaptation to the augmented matrix for efficient fine-tuning, we reduce the parameter burden during the fine-tuning process and enhance the model's generalization capability. Experimental results demonstrate that GraphSASA outperforms baseline methods.


\normalem
\bibliographystyle{IEEEtran}
\bibliography{refs.bib}
\end{document}